\documentstyle[12pt]{article}
\newcommand{\be}{\begin{equation}}
\newcommand{\ee}{\end{equation}}
\newcommand{\rar}{\rightarrow}

\begin{document}
 
\begin{titlepage}
\begin{flushright}
       {\bf UK/99--06}  \\
May 1999   \\
       hep-lat/9905033 \\
\end{flushright}
\begin{center}
 
{\bf {\LARGE A Noisy Monte Carlo Algorithm}}
 
\vspace{1cm}
 
{\bf L.\ Lin\footnote{On leave from Department of Physics, National Chung Hsing 
University, Taichung 40227, \mbox{Taiwan,} ROC},
 K.F.\ Liu, and
J.\ Sloan\footnote{Present address: Spatial Technologies, Boulder, CO}
} \\
[0.5em]
 {\it  Dept. of Physics and Astronomy, Univ. of Kentucky,
            Lexington, KY 40506}
       
\end{center}

\vspace{0.4cm}
 
\begin{abstract}

 We propose a Monte Carlo algorithm to promote Kennedy and Kuti's 
linear accept/reject algorithm which accommodates unbiased stochastic 
estimates of the probability to an exact one. This is achieved by adopting 
the Metropolis accept/reject steps for both 
the dynamical and noise configurations. We test it on the
five state model and obtain desirable results even for the case with large
noise. We also discuss its application to lattice QCD with stochastically
estimated fermion determinants.

\bigskip
 
 
\end{abstract}
 
\vfill
 
\end{titlepage}
 
\section{Introduction}                 \label{s1}
 
   Usually Monte Carlo algorithms require exact evaluation of the 
probability ratios in the accept/reject step. However, there are problems in
physics which involve extensive quantities such as the fermion determinant
which require $V^3$ steps to compute exactly. Thus the usual Monte Carlo
algorithms for a large volume are not numerically applicable to such problems 
directly. To address this problem, Kennedy and Kuti \cite{KK} proposed a Monte 
Carlo algorithm which admits stochastically estimated transition probabilities 
as long as they are unbiased. This opens up the door to tackling problems 
when it is feasible to estimate the transition probabilities but intractable
or impractical to calculate them exactly.
 
 The acceptance probability (denoted as $P_a$ from now on)
 in Kennedy-Kuti's linear algorithm is
\begin{equation} \label{lin}
  {\rm P}_a(U_1\rar U_2) = \left\{ 
\begin{array}{ll}
  \lambda^+ \, +\,\lambda^-\,
 \langle e^{\Delta H}\rangle\,\,\,\, ,
 & {\rm if}\,\,  f(U_1) > f(U_2)\,\, ,  \\
  \lambda^- \, +\,\lambda^+\, \langle e^{\Delta H}\rangle\,\,\,\, ,
 & {\rm if}\,\,  f(U_1)\le f(U_2)
\end{array} 
\right.
\end{equation}
where $\lambda^{\pm}$ are tunable real parameters ranging from 0 to 1,
$\langle e^{\Delta H}\rangle$ denotes an unbiased
estimator of the transition probability $e^{\Delta H}$ with 
$\Delta H = H(U_1)-H(U_2)$. $U_1$ denotes the old
configuration and $U_2$ the new or proposed configuration.
$f(U)$ is an observable of the configuration $U$ adopted for ordering
between $U_1$ and $U_2$.
Detailed balance can be proven to be satisfied~\cite{KK}.
 
But there is a drawback with this linear algorithm. The probability
$P_a$ could lie outside the interval between $0$
and $1$ since it is estimated stochastically. Once
the probability bound is violated, detailed balance
is lost and systematic bias will show up. It is hoped that if the
bound violation occurs rarely (e.g. once every million updates),
the systematic bias might be small enough so that the expectation values
of various quantities can still be correct within statistical errors \cite{KK}.
 
Within the framework of the linear algorithm, there are at least three
ways to reduce the probability of bound violation.
\begin{enumerate}
\item 
 The choice of $\lambda^+ = 0.0\,\, ,\,\, \lambda^-= 1/2$ in ref. \cite{KK}
can be generalized to 
\be  \label{eps}
\lambda^+ = 0.0\,\, ,\,\, \lambda^-={1\over 1+\alpha}\,\,\,.
\ee

With $\alpha > 1$, the allowed range
of $\langle e^{\Delta H}\rangle$ is proportionally increased so that the 
probability of upper bound violation can be
tamed, albeit at the expense of a lower acceptance rate.
 
\item
 One can choose a better ordering criterion to reduce the
bound violation. When the ordering criterion is
not correlated with $\Delta H$, the upper bound is violated more frequently 
than the case in which the $\Delta H$ itself is
used as the ordering criterion  (i.e. $\Delta H\ge$ or $< 0$). 
However, one cannot calculate $\Delta H$ exactly -- a premise for the
problem; the best one can do is to
estimate $\Delta H$ stochastically without bias. As long as it
can be made reasonably close to the true value
of $\Delta H$, it can be used as the ordering criterion. This should
greatly reduce the probability of upper-bound violation.

\item 
Usually it is $\Delta H$ that can be estimated without bias. Simple
exponentiation of this estimator, i.e. $e^{\langle \Delta H \rangle}$ 
inevitably yields a bias. However, it is demonstrated by Bhanot 
and Kennedy~\cite{BK} that an unbiased estimator 
$\langle e^{\Delta H}\rangle$ can be constructed via a series expansion
of the exponential in terms of the powers of independent unbiased estimator 
$\langle \Delta H\rangle$. One can reduce the variance of the estimated
acceptance probability by considering the variants of this series 
expansion. This will help reduce
the probability of both the lower-bound and upper-bound violations.
We tried several variants, the best turns out to be
\begin{equation}  \label{1/N}
\langle e^{\Delta H}\rangle\,\,\equiv\,\,\Pi_{i=1}^N\, e^{x_i},
\end{equation}
where $x_1, x_2, ...., x_N$ are identical,
independent unbiased estimators of ${\Delta H\over N}$,
and each $e^x$ is estimated by the series expansion
developed by Bhanot and Kennedy~\cite{BK}:
\be \label{bk}
\langle {e^x}\rangle\, = \, 1\,+\, x_1(1\, +\, {1\over 2}x_2(1\, +\, 
{1\over 3}x_3 (1\, + .... )))\,\,
\ee
where the coefficients in the Taylor expansion are interpreted as
probabilities. The procedure goes as follows. First, one sets 
$\langle {e^x}\rangle=1+x_1$. Then one adds $x_1 x_2$
to $\langle {e^x}\rangle$ with probability ${1\over 2}$; otherwise one stops.
If it is not stopped, one then continues to add $x_1 x_2 x_3$ to 
$\langle {e^x}\rangle$ with probability ${1\over 3}$, and so on.
It is easy to prove~\cite{BK} that the above estimator is unbiased.
One can also calculate its variance which is
\begin{equation}   \label{var}
Var(\langle e^{\Delta H}\rangle) = \{e^{\frac{\overline{\Delta H}^2 + \delta^2}
{N^2}} + 2(e^{\frac{\overline{\Delta H}}{N}} - e^{\frac{\overline{\Delta H}^2 
+ \delta^2}{N^2}}) \frac{\overline{\Delta H}/N}{\overline{\Delta H}/N -
(\overline{\Delta H}^2 + \delta^2)/N^2}\}^N - e^{2\overline{\Delta H}}
\end{equation} 
where $\delta^2 = \overline{\Delta H^2} - \overline{\Delta H}^2$ is the
variance of ${\Delta H}$ from the noise estimate. It is smaller than
those of the other variants of the series expansion we considered such
as $e^{\frac{1}{N}\sum_{i =1}^N x_i}$ and $\frac{1}{N} \sum_{i= 1}^N
e^{x_i}$. It can be shown that only a finite number of terms are needed 
in actual calculations.
\end{enumerate}

Although one can improve the performance of the linear algorithm with
the above techniques, there are
still problems inherent to the algorithm which are impossible to eradicate.
First of all, if we assume that the estimator of the acceptance probability
has a Gaussian distribution, then the long 
tails of the Gaussian distribution always exist. As a result, the probability 
bound violations will never be completely excluded.
Secondly, the linear algorithm with a stochastic estimator
is a volume-squared algorithm. Thus, in realistic simulations of problems
such as lattice QCD with dynamical fermions, it would be very costly to 
estimate the fermion determinant with sufficient accuracy in order to put 
the probability bound violations under control. The volume dependence 
can be seen from the following consideration. Suppose we take the best 
estimator in Eq. (\ref{1/N}) to calculate the acceptance probability,
the variance is primarily a function of $\delta^2/N$ when $\overline{\Delta H}
\simeq 0$, according to Eq. (\ref{var}). If in addition $\delta^2/N$ is 
small, one can do an expansion and find that the variance goes like 
$\delta^2/N$. Usually $\delta^2$ is proportional to the
volume or the size of the problem. Therefore, if one wants to keep 
the bound violations the same as the volume grows, one needs to have a 
larger $N$. Consequently, $N$ grows as the volume $V$ and the cost 
will be proportional to $V^2$ since the cost of the stochastic estimator 
itself is usually proportional to $V$, e.g. for a sparse matrix.

In order to completely
remove any systematic bias coming from probability bound violations
and to reduce the cost of simulation
on large volumes, one needs to go beyond the linear
algorithm. In this letter, we propose
a new algorithm which will achieve these goals. We shall see that
the new algorithm
eliminates the upper bound violation and absorbs the negative sign of the
lower bound violation into the observable. 
These are achieved by introducing auxiliary variables and
going back to the Metropolis accept/reject criterion.

\section{A Noisy Monte Carlo Algorithm}

Let us consider a model with Hamiltonian $H(U)$ where $U$ collectively 
denotes the dynamical variables of the system.
The major ingredient of the new approach is
to transform the noise for the stochastic estimator into stochastic variables.
To this end, the stochastic series expansion in Eq. (\ref{bk}) is written 
in terms of an integral of the stochastic variables $\xi$. 

\be \label{fU}
e^{-H(U)}\,\, =\,\,\int [D\xi]\,P_\xi(\xi)\,\, f(U,\xi),
\ee
where $f(U,\xi)$ is an unbiased estimator of $e^{-H(U)}$ from the 
stochastic variable $\xi$ and $P_\xi$ is the probability distribution 
for $\xi$.
Here, we use $\xi$ as a collective symbol for all the stochastic variables.

Given this integral in Eq. (\ref{fU}), the partion function of the model
can be written as
\begin{eqnarray}  \label{Z}
Z &=&  \int [DU]\, e^{-H(U)} \nonumber \\ 
  &=& \int [DU][D\xi]P_\xi(\xi)\, f(U,\xi).
\end{eqnarray}
 
Originally, we have a configuration space of $U$. Now it is enlarged to
$(U,\xi)$ with the inclusion of the stochastic variable $\xi$.
From now on, we shall specify a configuration or state in this enlarged space.
 
 
The next step is to address the lower probability-bound violation. One
first observes that
\begin{equation}
f(U,\xi)\,\, = \,\,{\rm sign}(f)\,|f(U,\xi)|\,\, .
\end{equation}
Since ${\rm sign}(f)$, the sign of $f$,
is a state function, we can write the expectation value
of the observable $O$ as
\begin{equation}  \label{O}
\langle O \rangle = \int[DU][D\xi]\,P_\xi(\xi)
  \,O(U)\,{\rm sign}(f)\,|f(U,\xi)|/Z. 
\end{equation}

After redefining the partition function to be
\begin{equation}
 Z = \int [DU][D\xi]P_\xi(\xi)\, |f(U,\xi)|,
\end{equation}
which is semi-positive definite, 
the expectation of $O$ in Eq. (\ref{O}) can be rewritten as
\begin{equation}  \label{Onew}
\langle O \rangle = \langle O(U) \,{\rm sign}(f) \rangle/\langle 
 {\rm sign}(f) \rangle.
\end{equation} 
As we see, the sign of $f(U,\xi)$ is not a part of the probability any more
but a part in the observable.
Notice that this reinterpretation is possible because the sign of
$f(U,\xi)$ is a state function which depends on the configuration of $U$ and
$\xi$. We note that in the earlier linear accept/reject case 
in Eq. (\ref{lin}),
the acceptance criterion depends on the transition probability 
$\langle e^{\Delta H}
\rangle = \langle e^{H(U_1) - H(U_2)}\rangle$ which cannot be factorized into
a ratio of state functions such
as $\langle e^{H(U_1)}\rangle/\langle e^{H(U_2)}\rangle$.
Consequently, the sign of the acceptance probability in the linear 
algorithm~\cite{KK} cannot be swept into the observable as in Eq. (\ref{Onew}).

It is clear then, to avoid the problem of lower probability-bound 
violation, the accept/reject criterion has to be factorizable into
a ratio of the new and old probabilities so that the sign of the
estimated $f(U,\xi)$ can be absorbed into the observable.  
This leads us back to the Metropolis accept/reject criterion which
incidentally cures the problem of upper probability-bound violation at
the same time. It turns out two accept/reject steps are needed in general.
The first one is to propose updating of $U$ via some procedure
while keeping the stochastic variables $\xi$ fixed.
The acceptance probability $P_a$ is
\be  \label{met1}
 P_a(U_1,\xi \rightarrow U_2,\xi)\,\,= \,\,
  {\rm min}\Bigl(1,{|f(U_2,\xi)|\over
                    |f(U_1,\xi)|}\Bigr)\,\,  .
\ee
The second accept/reject step involves the refreshing of the stochastic 
variables $\xi$ according to the probability distribution $P_{\xi}(\xi)$ while 
keeping $U$ fixed. The acceptance probability is 
\be   \label{met2}
  P_a(U,\xi_1\rightarrow U,\xi_2)\,\,= \,\,
  {\rm min}\Bigl(1,{|f(U,\xi_2)|\over |f(U,\xi_1)|}\Bigr)\,\,  .
\ee
It is obvious that there is neither lower nor upper probability-bound 
violation in either of these two Metropolis accept/reject steps. 
Furthermore, it involves the ratios of separate state functions so that
the sign of the stochastically estimated probability $f(U,\xi)$ can be 
absorbed into the observable as in Eq. (\ref{Onew}).  
 
Detailed balance can be proven to be satisfied.
For the first step which involves the updating $U_1\rar U_2$
with $\xi = $ fixed, one can show for the case
$|f(U_2,\xi)|/ |f(U_1,\xi)| < 1$
\begin{eqnarray} \label{db1}
&&P_{eq}(U_1,\xi)\, P_c(U_1\rar U_2)
\, P_a(U_1,\xi\rar U_2,\xi) \nonumber \\
&-&\,P_{eq}(U_2,\xi)\, P_c(U_2\rar U_1)
\, P_a(U_2,\xi\rar U_1,,\xi) \nonumber \\
&=&P_\xi(\xi)\,|f(U_1,\xi)|\, P_c(U_1\rar U_2)\,
{|f(U_2,\xi)|\over |f(U_1,\xi)|} \nonumber \\
&-& P_\xi(\xi)\,|f(U_2,\xi)|\, P_c(U_2\rar U_1)
\nonumber \\
&=&P_\xi(\xi)\, |f(U_2,\xi)|\,  P_c(U_1\rar U_2)\nonumber \\
&-&P_\xi(\xi)\, |f(U_2,\xi)|\,  P_c(U_2\rar U_1)\,
=\, 0\,\, .
\end{eqnarray}
where $P_{eq}$ is the equilibrium distribution and $P_c$ is the
probability of choosing a candidate phase space configuration satisfying 
the reversibility condition 
\be
 P_c(U_1\rar U_2)\,=\, P_c(U_2\rar U_1)\,\, .
\ee
Detailed balance for the second step which invokes the updating
$\xi_1\rar\xi_2$ with $U$ fixed
can be similarly proved. For the case
$ |f(U_2,\xi)|/ |f(U_1,\xi)|\,<\, 1$, we have
\begin{eqnarray}
&&P_{eq}(U,\xi_1)\, P_c(\xi_1\rar\xi_2)\,
P_a(U,\xi_1\rar U,\xi_2) \nonumber \\
&-&P_{eq}(U,\xi_2)\, P_c(\xi_2\rar\xi_1)\,
P_a(U,\xi_2\rar U,\xi_1) \nonumber \\
&=&\,P_\xi(\xi_1)\,|f(U,\xi_1)|\, P_\xi(\xi_2)\,
{|f(U,\xi_2)|\over |f(U,\xi_1)|}
-\,P_\xi(\xi_2)\,|f(U,\xi_2)|\, P_\xi(\xi_1) \nonumber \\
&=& 0.
\end{eqnarray}
Therefore, this new algorithm does preserve detailed balance and is 
completely unbiased.

We have tested this noisy Monte Carlo (NMC) on a 5-state model
which is the same used in the linear algorithm~\cite{KK} for demonstration.
Here, $P_c(U_1\rar U_2) = {1\over 5}$ and we use Gaussian noise to mimic
the effects of the noise in the linear algorithm and the stochastic
variables $\xi$
in NMC. We calculate the average energy with the linear algorithm and 
the NMC. Some data are presented in Table 1. Each data point is obtained
with a sample of one million configurations. The exact value
for the average energy is $0.180086$.

\begin{table}[tb]
\caption{  \label{t1} Data for the average energy 
obtained by NMC and the linear algorithm~\cite{KK}. They are obtained with
a sample size of one million configurations each. $Var$ is the variance
of the noise estimator for $e^{-H}$ in NMC and $e^{\Delta H}$ in the 
linear algorithm. $\alpha$ in Eq. (\ref{eps}) is set to 1.0 in the 
latter case. Negative Sign denotes the percentage of times when the sign of
the estimated probability is negative in NMC.
Low/High Vio. denotes the percentage of times 
when the low/high probability-bound is violated in the linear algorithm.
The exact average energy is $0.180086$.}
\begin{center}
\begin{tabular}
{|c|c|c||c|c|c|}
\hline
{$Var$}&NMC&Negative Sign&Linear &Low Vio.& High Vio.
\\
\hline
0.001 &  0.17994(14)  &  0\%  & 0.18024(14)  &    0\% &     0\%\\
\hline
0.002 &  0.18016(14)  &  0\%  & 0.17994(14)  &    0\% &     0\%\\
\hline
0.005 &  0.18017(14)  &  0\%  & 0.17985(14)  &    0\% &     0\%\\
\hline
0.008 &  0.17993(14)  &  0\%  & 0.17997(14)  &    0\% &     0\%\\
\hline
0.01  &  0.18008(14)  &  0\%  & 0.17991(14)  &    0\% &     0\%\\
\hline
0.06  &  0.17992(14)  &  0.008\%& 0.17984(14)&  0.001\%&   0.007\%\\
\hline
0.1   &  0.17989(14)  &   0.1\%&  0.17964(14)&   0.1\% &    0.3\%\\
\hline
0.2   &  0.18015(15)  &   1.6\%&  0.18110(13)&    1\%&       1\%\\
\hline
0.5   &  0.1800(3)  &    5\% &    0.1829(1)&    3\%&       4\%\\
\hline
1.0   &  0.1798(4)  &  12\%&     0.1860(1)&     6\%&       7\%\\
\hline
5.0   &  0.1795(6)  &  28\%&     0.1931(1)&    13\%&      13\%\\
\hline
6.5   &  0.1801(5)  &  30\%&     0.1933(1)&    13\%&      14\%\\
\hline
10.0  &  0.1799(9)  &  38\%&      -       &     - &       - \\            
\hline
15.0  &  0.1798(9)  &  38\%&      -       &     - &       - \\            
\hline
20.0  &  0.1803(11)  &  39\%&     -       &     - &       - \\            
\hline
30.0  &  0.1800(13)  &  41\%&     -       &     - &       - \\            
\hline
50.0  &  0.1794(17)  &  44\%&     -       &     - &       - \\            
\hline
\end{tabular}
\end{center}
\end{table}

We first note that as long as the variance of the noise is less than 
or equal to 0.06, the statistical errors of both the NMC and 
linear algorithm stay the same and the results are correct within 
two $\sigma$. To the extend that the majority of the numerical effort in a 
model is spent in the stochastic estimation of the probability,
this admits the possibility of a good deal of saving through the reduction
of the number of noise configurations, since a poorer estimate of the 
probability with 
less accuracy works just as well. As the variance becomes larger than 0.06,
the systematic bias of the linear algorithm sets in and eventually becomes
intolerable, while there is no systematic bias in the NMC results. 
In fact, we observe that the NMC result is correct even 
when the percentage of negative probability reaches as high as 44\%, 
although the statistical fluctuation becomes larger due to the fact that
the negative sign appears more frequently. We should remark that the
Metropolis acceptance rate is about 92\% for the smallest noise variance.
It decreases to 85\% when the varicance is 0.1 and it drops eventually
to 78\% for the largest variance 50.0. Thus, there is no serious degrading 
in the acceptance rate when the variance of the noise increases.

We further observe that the variance of the NMC result does not grow as 
fast as the variance of the noise. For example, the variance of the 
noise changes by a factor of 833 from 0.06, where the
probability-bound violation starts to show up in the linear algorithm, to
50.0. But the variance of the NMC result is only increased by a factor
of $(0.0017/0.00014)^2 = 147$. Thus, if one wants to use the linear algorithm
to reach the same result as that of NMC and restricts to configurations 
without probability-bound violations, it would need 833 times the noise 
configurations to perform the stochastic estimation in order to bring the 
noise variance 
from 50.0 down to 0.06 but 147 times less statistics in the Monte Carlo 
sample. In the case where the majority of the computer time is consumed in
the stochastic estimation, it appears that NMC can be more economical than the
linear algorithm. 

 
\section{Lattice QCD with Fermion Determinant}

One immediate application of NMC is lattice QCD with dynamical fermions.
The action is composed of two parts -- the pure gauge action
$S_g(U)$ and a fermion action $S_F(U) = - Tr \ln M(U)$.
Both are functionals of the gauge link variables $U$.  
Considering the Hybrid Monte Carlo \cite{HBMC} approach with 
explicit $Tr \ln M$
for the fermion action,  we first enlarge the phase space from 
$(U)$ to $(U,p)$ where $p$ denotes the conjugate momentum of $U$.
The partion function is 
\begin{equation}
Z =  \int [DU][Dp]\, e^{-H(U,p)}
\end{equation}
where $H(U,p)=p^2/2+S_g(U)+S_F(U)$.
To apply NMC, we introduce stocahstic variables to estimate 
the fermion determinant
\begin{equation}  \label{det}
det M(U) = e^{Tr \ln M} = \int [D\xi] P_\xi(\xi)\,  f(U,\xi)
\end{equation}
where $f(U,\xi)$ will be given later.
The partition function is then
\begin{equation}  \label{Z2}
Z = \int [DU][Dp][D\xi]P_\xi(\xi)\, e^{-H_G(U,p)}\, f(U,\xi),
\end{equation}
where $H_G = p^2/2+S_g$.
In the Hybrid Monte Carlo, the configuration $(U,p)$ is updated 
with molecular dynamics. In this case, the probability of choosing
a candidate configuration is $P_c(U_1,\,p_1 \rightarrow U_2,\,p_2) =
\delta[(U_2,\,p_2) - (U_1(\tau),\,p_1(\tau))]$ where $U_1(\tau)$ and
$p_1(\tau)$ are the evolved values at the end of the molecular dynamics
trajectory after $\tau$ steps. Using the reversibility condition
\begin{equation}
P_c(U_1,\,p_1 \rightarrow U_2,\,p_2) = P_c(U_2,\, -p_2 \rightarrow
U_1,\, -p_1),
\end{equation} 
one can again prove detailed balance with two corresponding Metropolis
steps as in Eqs. (\ref{met1}) and (\ref{met2}).  

To find out the explicit form of $f(U,\xi)$, we
note that the fermion determinant can be calculated stochastically as
a random walk process~\cite{BK}
\begin{equation}
e^{Tr\ln M} = 1 + Tr\ln M (1 + \frac{Tr \ln M}{2} (1 + \frac{Tr \ln M}{3}
(...))) \,\, ,
\end{equation}
as described in Eq. (\ref{bk}). This can be expressed in 
the following integral
\begin{eqnarray}  \label{trln}
&& e^{Tr\ln M}=\int \prod_{i =1}^{\infty}  d\,\eta_i \,
P_{\eta}(\eta_i)
     \int_{0}^1 \prod_{n =2}^{\infty}  d\,\rho_n  \nonumber \\
&& [1 + \eta_1^{\dagger} \ln M \eta_1
  (1 + \theta(\rho_2 - {1\over 2}) \eta_2^{\dagger} \ln M \eta_2
  (1 + \theta(\rho_3 - {2\over 3}) \eta_3^{\dagger} \ln M \eta_3 (...],
\end{eqnarray}
where $P_{\eta}(\eta_i)$ is the probability distribution for the
stochastic variable $\eta_i$. It can be the Gaussian noise or the $Z_2$ noise 
($P_{\eta}(\eta_i) = \delta(|\eta_i| -1)$ in this case). The latter is
preferred since it has the minimum variance~\cite{dl94}. $\rho_n$ is
a stochastic variable with uniform distribution between 0 and 1.
This sequence terminates stochastically in finite time and only the
seeds from the pseudo-random number generator need to be stored in practice.
Comparing this to Eq. (\ref{det}), the function $f(U,\eta,\rho)$ (Note 
the $\xi$ in Eq. (\ref{det}) is represented by two stochastic variables 
$\eta$ and $\rho$ here) is represented by the part of the integrand 
between the the square brackets in Eq. (\ref{trln}). 
One can then use the efficient 
Pad\'{e}-Z$_2$ algorithm~\cite{TDLY} to calculate the $\eta_i\ln M \eta_i$ in
Eq. (\ref{trln}). All the techniques for reducing the
variance of the estimator without bias developed before~\cite{TDLY} 
can be applied here. It is learned that after the unbiased subtraction,
the error on the stochastic estimation of the $Tr \ln M$ difference 
between two fermion matrices at the beginning and end of the molecular
dynamics trajectory for an $8^3 \times
12$ lattice with Wilson fermion at $\beta = 6.0$ and $\kappa = 0.154$ can be
reduced from 12.0 down to 0.49 with a mere 100 noise configurations~\cite
{TDLY}. This implies a 49\% error on the determinant ratio and makes the
application of the noisy Monte Carlo algorithm rather promising.

Finally, there is a practical concern that $Tr\ln M$ can be large
so that it takes a large statistics to have a reliable estimate of
$e^{Tr\ln M}$ from the series expansion in Eq. (\ref{trln}). In general, 
for the Taylor expansion $e^x = \sum x^n/n!$, the series will start to
converge when $x^n/n! > x^{n + 1}/(n + 1)!$. This happens at $n = x$.
For the case $x = 100$, this implies that one needs to have more than 100! 
stochastic configurations in the Monte Carlo
integration in Eq. (\ref{trln}) in order to have a convergent estimate.
Even then, the error bar will be very large. To avoid this difficulty, one
can implement the following strategy. First one note that since the 
Metropolis accept/reject involves the ratio of exponentials, one can subtract 
a universal number $x_0$ from the exponent $x$ in the Taylor expansion 
without affecting the ratio. Second one can use the trick in Eq. (\ref{1/N})
to diminish the value of the exponent. In other words, one can replace $e^x$ 
with $(e^{(x - x_0)/N})^N$ to satisfy $|x - x_0|/N < 1$. The best choice for
$x_0$ is $\overline{x}$, the mean of $x$. In this case, the variance 
in Eq. (\ref{var}) becomes $e^{\delta^2/N} -1$. Comparing with Eq. (\ref{var}),
one can verify that it is smaller than the case
without $\overline{x}$ subtraction by $e^{2\overline{x}}$.  
We should mention that this is not an issue in the Kennedy-Kuti
algorithm where the accept/reject criterion involves the 
transition probability   	
$e^{\Delta H} = e^{H(U_1) - H(U_2)}$ not the ratio of probabilities
as in the Metropolis criterion.
 
\section{Summary and Discussion} \label{s5}
 
In summary, the new noisy Monte Carlo algorithm proposed here
is free from the problem of probability-bound
violations which afflicts the linear accept/reject algorithm,
especially when the variance of the noise is large. 
The upper-bound violation is avoided by going back to the Metropolis
accept/reject. The lower-bound violation problem is
tackled by grouping the sign of the estimated probability with the observable.
With the probability-bound violation problem solved, NMC is a bona fide
unbiased stochastic algorithm as demonstrated in the 5-state model.
Furthermore, it is shown in the 5-state model that it is not necessary to 
have an extremely small variance in the stochastic estimation. 
With the encouraging results from the Pad\'{e}-Z$_2$ estimation of the
$Tr \ln M$~\cite{TDLY}, one has a reasonable hope that the $V^2$ dependence 
of NMC will be tamed with a smaller prefactor. We will apply NMC to the 
dynamical fermion updating in
QCD and compare it to the HMC with pseudo-fermions \cite{FUT}.

\section{Acknowledgment}
 
The authors would like to thank A. Kennedy for stimulating discussions and
for helping clarify the Metropolis steps.
This work is partially supported by the U.S. DOE grant DE-FG05-84ER40154.
Lee Lin wishes to thank NSC of the Republic of China
for support under the contract number 36083F. He is also grateful to members
of Department of Physics and Astronomy of University of Kentucky for the
hospitality extended to him during his stay in Lexington, Kentucky.
Keh-Fei Liu wishes to thank the National Center for Theoretical Sciences,
Hsinchu, Taiwan for the invitation to visit their Center where this work
is finished. 

\end{document}